\def\year{2020}\relax
\documentclass[letterpaper]{article} 
\usepackage{aaai20}  
\usepackage{times}  
\usepackage{helvet} 
\usepackage{courier}  
\usepackage[hyphens]{url}  
\usepackage{graphicx} 
\usepackage{graphicx}  
\newcommand{\citet}[1]{\citeauthor{#1} \shortcite{#1}}

\title{Prevalence of Low-Credibility Information on Twitter \\ During the COVID-19 Outbreak}

\author{
Kai-Cheng Yang, Christopher Torres-Lugo, Filippo Menczer\\
Observatory on Social Media, Indiana University, Bloomington, IN, USA 
}

\begin{document}
\maketitle

\begin{abstract}
As the novel coronavirus spreads across the world, concerns regarding the spreading of misinformation about it are also growing.
Here we estimate the prevalence of links to low-credibility information on Twitter during the outbreak, and the role of bots in spreading these links.
We find that the combined volume of tweets linking to low-credibility information is comparable to the volume of \textit{New York Times} articles and CDC links.
Content analysis reveals a politicization of the pandemic. The majority of this content spreads via retweets. 
Social bots are involved in both posting and amplifying low-credibility information, although the majority of volume is generated by likely humans. Some of these accounts appear to amplify low-credibility sources in a coordinated fashion. 
\end{abstract}

\section{Introduction}

Today most countries are experiencing an unprecedented outbreak of the novel coronavirus (COVID-19). 
Millions of people have tested positive for the virus and tens of thousands people have died from it globally (\url{coronavirus.jhu.edu/map.html}).
At the same time, we have observed an increase of approximately 25\% in Twitter volume. 

With millions of people stuck in their homes and accessing information via social media, concerns about the spread of misinformation about the pandemic (referred to as ``infodemic''~\cite{zarocostas2020fight}) have mounted. 
Social media facilitate the spread of misinformation~\cite{vosoughi2018spread}, manipulation~\cite{stella2018bots}, and radicalization of users~\cite{thompson2011radicalization}.
These issues are even more pressing in the current atmosphere since the information flowing through social media is directly related to the public health and safety.

In response, quite a few research papers have been made public lately that estimate the prevalence of COVID19-related misinformation on social media~\cite{cinelli2020covid,pulido2020covid,laato2020people} and characterize the behaviors of inauthentic actors~\cite{gallotti2020assessing,ferrara2020covid}.
These studies use different methods on different datasets and yield different perspectives on the issue.
However, given the complex nature of the problem, many questions remain unanswered.
In this paper, we use a random sample of tweets to estimate the prevalence of COVID19-related low-credibility information on Twitter and further characterize the role of social bots~\cite{ferrara2016rise,shao2018spread}.

\section{Methods}

\subsection{Identification of low-credibility information}
\label{sec:low_credibility}

Identification of misinformation often requires fact-checking from experts, which is extremely time consuming and therefore not viable for this analysis. 
Instead, we focus on the URLs embedded in the tweets and annotate the credibility of the content not at the URL level but at the domain level, following the literature~\cite{Lazer-fake-news-2018}.

As there is no consensus on the definition of misinformation, we focus on a broad set of low-credibility (including hyperpartisan) sources. 
Our list comes from recent research and includes sources that fulfill any one of the following criteria: (1)~labeled as low-credibility by \citet{shao2018spread}; (2)~labeled as ``Black'' or ``Red'' or ``Satire'' by \citet{grinberg2019fake}; (3)~labeled as ``fakenews'' or ``hyperpartisan'' by \citet{pennycook2019fighting}; or (4)~labeled as ``extremeleft'' or ``extremeright'' or ``fakenews'' by \citet{bovet2019influence}.
This gives us a list of 570 low-credibility sources.

\subsection{Data collection}

We collected two datasets using different methods. 

\texttt{DS1} consists of tweets containing a set of hashtags and links.
Various hashtags are associated to the coronavirus~\cite{chen2020covid}, but some are focused on certain aspects of the outbreak and some reflect certain biases. 
To provide a general and unbiased view of the discussion, we chose two generic hashtags \texttt{\#coronavirus} and \texttt{\#covid19} as our seeds.
Our data was collected using an API from the Observatory on Social Media, which allows to search tweets from a 10\% random sample of public tweets~\cite{davis2016osome}.
This dataset consists of tweets from 
Mar. 9--29, 2020.

Estimating the prevalence of low-credibility information requires matching the URL domains from the tweets against the list defined above.
To include all links from the tweet objects obtained through the API, we used a regular expression to extract any URL-like strings from the tweet text in addition to fetching URLs from the entity metadata.
For retweets, we also included the URLs in the original tweets using the same method.

Since shortened URLs are very common, we identified those from 70 most frequent shortening services and expanded the URLs through HTTP requests to obtain the real domains.
The expanded URLs were further cleaned.
We removed those linking to Twitter itself, which turn out to be the majority, and those linking to other social media sites. 
While these links might still lead to low-credibility information, there is no easy way to verify, so we exclude them in this study.
The remaining links mainly belong to news outlets and authorities like government agencies.
The \texttt{DS1} is the set of tweets that match the COVID-19 hashtags \emph{and} containing any of these links.

\texttt{DS2} starts from a collection of tweets containing links to low-credibility sources. 
The data was collected using the Twitter streaming/filter API from Feb. 1 to Apr. 27, 2020.
The URLs were extracted and the corresponding web pages were fetched.
To reveal common low-credibility information topics, we analyzed the titles of the linked articles and retained those with keywords ``coronavirus'' and ``covid.''
We ranked the links by the number of tweets containing them and extracted the top 1,200. 
Each URL in \texttt{DS2} has been shared at least 50 times. 

\subsection{Bot detection}

Social bots are social media accounts controlled in part by algorithms~\cite{ferrara2016rise}.
Malicious bots are known to spread low-credibility information~\cite{shao2018spread} as well as creating confusion in the online debate about health-related topics like vaccination~\cite{broniatowski2018weaponized}.
It is therefore interesting to characterize the role of social bots in spreading COVID19-related low-credibility information.

We adopt BotometerLite~\cite{yang2019scalable}, a bot detection model that enables large-scale bot detection. 
By strategically selecting a subset of the training dataset, BotometerLite achieves high accuracy in cross-validation as well as cross-domain tests. 
BotometerLite generates a score between 0 and 1 for every account, with higher scores indicating bot-like profiles.
For binary classification of accounts we use a threshold of 0.5 in this paper.

\section{Results}

\subsection{Prevalence of low-credibility information}

\begin{figure}[t]
    \centering
    \includegraphics[width=\columnwidth]{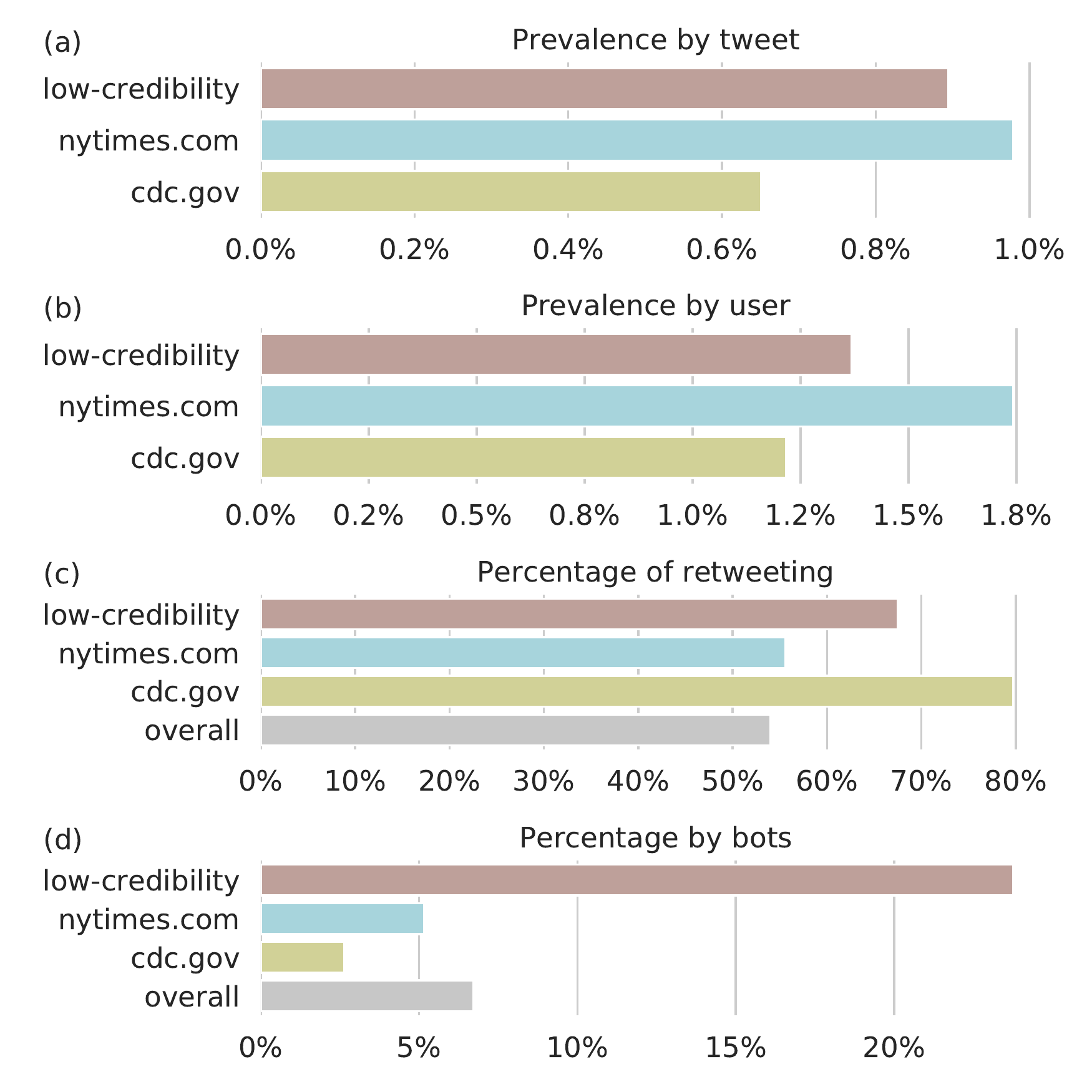}
    \caption{
    Prevalence of low-credibility, \textit{nytimes}, and \textit{cdc.gov} links in \texttt{DS1}, at the level of (a)~tweets and (b)~users. Percentage of (c)~retweets and (d)~tweets by likely bots for low-credibility, \textit{nytimes}, \textit{cdc.gov}, and overall links in \texttt{DS1}.
    }
    \label{fig:prevalence}
\end{figure}

To report on the prevalence of low-credibility information, we obtain reference volume levels using links to the New York Times, a mainstream news source, and the CDC, an official source of critical information related to the outbreak. 
Our results show that links to low-credibility sources combined contribute 0.89\% of the total tweet volume in \texttt{DS1} (Fig.~\ref{fig:prevalence}(a)).
For comparison, \textit{nytimes.com} contributes 0.98\% and \textit{cdc.gov} contributes less than 0.65\%.
To account for the fact that some users might share certain information repeatedly, we provide the same analysis at the level of users, i.e., the percentage of users who shared the corresponding links at least once, in Fig.~\ref{fig:prevalence}(b).
The results are qualitatively similar.
These findings suggest that  low-credibility information is not rampant on Twitter, but it does have a volume share comparable with highly reliable sources.

We also show the percentage of retweets for different sources in Fig.~\ref{fig:prevalence}(c).
About 68\% of the links to low-credibility information are shared by retweets.
For comparison, this fraction is about 54\% for \textit{nytimes.com} and all URLs together in \texttt{DS1}. 
This suggests that users involved with low-credibility information on Twitter are more likely to share links posted by others.
Interestingly, \textit{cdc.gov} has an even hither retweet rate.

\begin{figure*}
    \centering
    \includegraphics[width=1\textwidth]{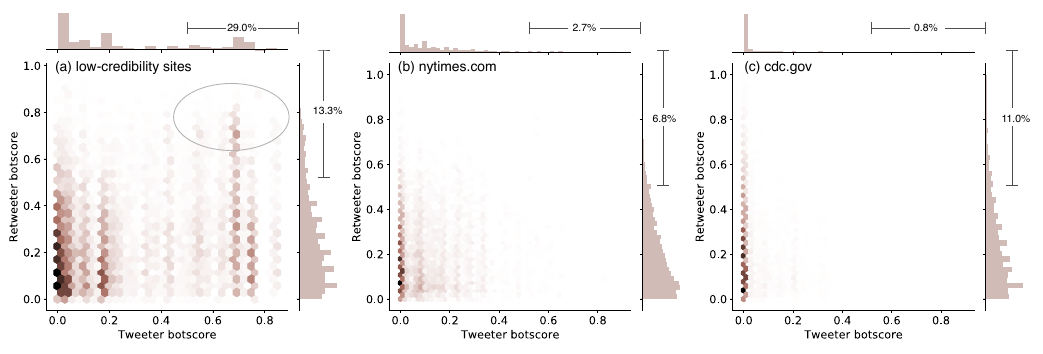}
    \caption{Joint distribution of bot scores of tweeters and retweeters for (a)~low-credibility information; (b)~\textit{nytimes.com}; and (c)~\textit{cdc.gov}.
    Annotations show the tweet volumes contributed by likely bots.}
    \label{fig:fake_bot_interaction}
\end{figure*}

\subsection{Role of social bots}

We report the percentage of tweets posted by social bots for different sources in Fig.~\ref{fig:prevalence}(d).
A significantly higher ratio of the volume of low-credibility information is shared by likely bot accounts, compared to the volume of tweets linking to reliable sources and the overall baseline. 
Since some accounts post multiple tweets with the same link, affecting the bot ratio estimation, we also perform the same analysis at the user level (not shown in Fig.~\ref{fig:prevalence}).
The bot ratios become 12.1\%, 6.5\%, 10.6\%, and 11.7\% for low-credibility, \textit{nytimes.com}, \texttt{cdc.gov}, and overall links, respectively.
The decreases in bot volume ratio compared to the tweet-level analysis suggest that low-credibility links are amplified by hyper-active bots. These results are consistent with previous findings~\cite{shao2018spread}.

Let us focus on retweets to characterize the interaction between original posters of links and accounts that amplify the reach of those links.
The results for different sources are shown in Fig.~\ref{fig:fake_bot_interaction}.
For low-credibility information, although the majority of tweets and retweets are posted by human-like accounts, we do see a higher-than-normal participation rate of bot accounts in both posting and amplifying the content.
We also find that bot-like tweeters attract more bot-like retweeters than human-like tweeters (see highlighted oval in Fig.~\ref{fig:fake_bot_interaction}(a)), despite a majority of human-like retweeters.

We see in Fig.~\ref{fig:fake_bot_interaction}(b) that bot participation is much lower for \textit{nytimes.com} than low-credibility sources.
The pattern for \textit{cdc.gov} is  interesting (Fig.~\ref{fig:fake_bot_interaction}(c)).
Original posting is dominated by the official account \texttt{@CDCgov}, whereas the retweeters have a much broader score spectrum with a relatively high proportion of bot-like accounts.
This suggests that some bots are also trying to disseminate useful information, in agreement with previous findings~\cite{ferrara2020covid}.

\subsection{Coordinated amplification of low-credibility information}

\begin{figure}[t]
    \centering
    \includegraphics[width=0.75\columnwidth]{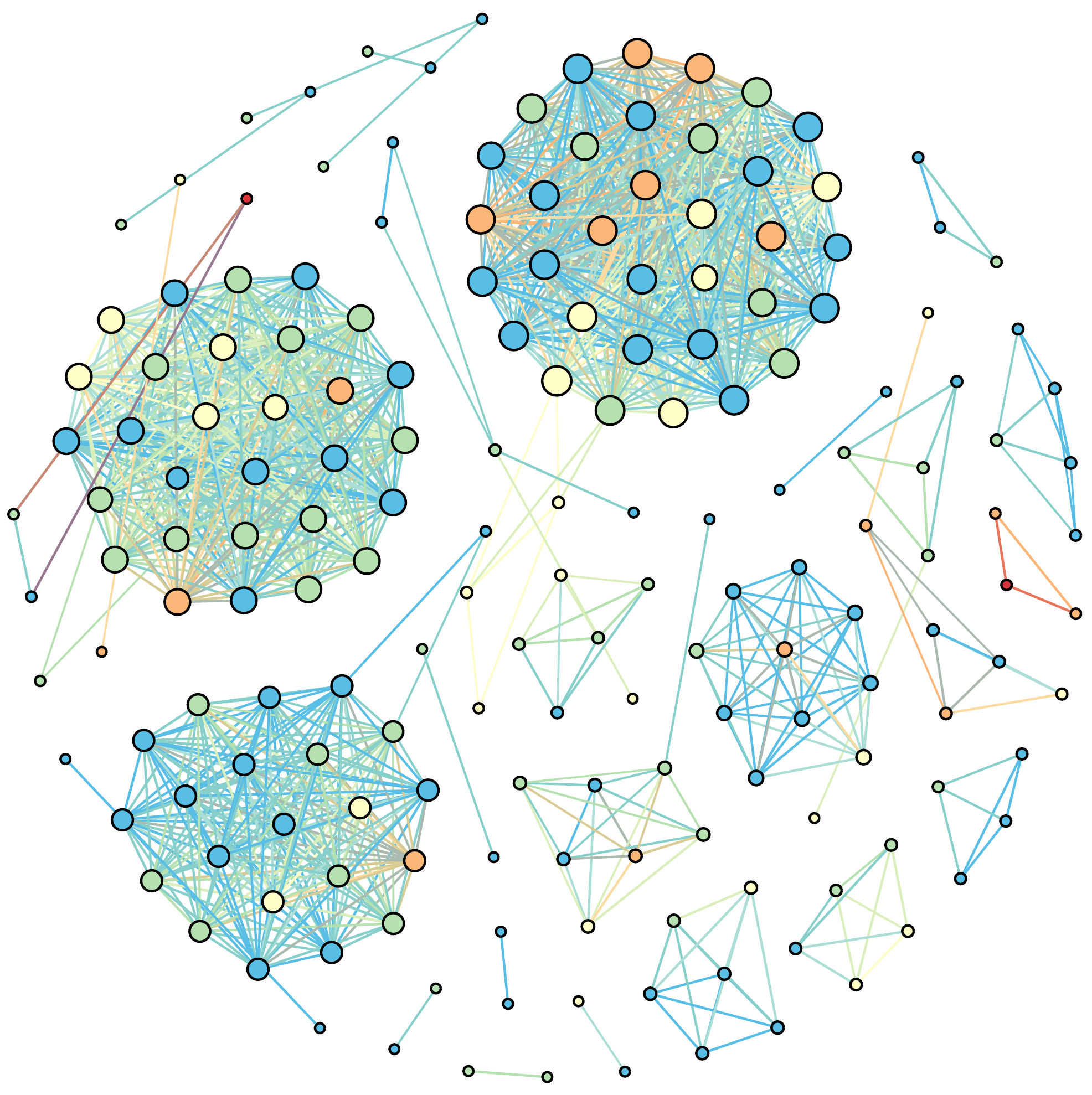}
    \caption{Similarity network of accounts sharing links to low-credibility sources. 
    Nodes are colored using a human-like (blue) to bot-like (red) scheme and size is proportional to strength (weighted degree). Edge weights represent cosine similarity among link vectors (see text).
    Only links with weight above 0.8 are shown, and singleton nodes after this filtering are removed. The final network consists of 180 nodes and 1,343 edges.
    }
    \label{fig:rt_net}
\end{figure}

Let us build a network of shared low-credibility domains to highlight potentially coordinated groups of accounts amplifying misinformation~\cite{pacheco2020uncovering}.
To build this network we assume characteristic behavior from coordinated accounts, in this case that they will work together to amplify misinformation from the same sources and at fairly similar, if not identical rate. Based on this assumption, we proceed to measure the similarity between pairs of accounts.
We focus on accounts that share at least 3 links. 
Then we extract domains from shared links to low-credibility sources, and represent an account as a vector of such domains. We finally calculate the cosine similarity between each pair of account vectors and use it as a network weight. 

The resulting network is shown in Fig.~\ref{fig:rt_net}. We note a few densely connected clusters of accounts, which share links to many of the same low-credibility sources. Although this network is not dominated by accounts with high bot scores, manual inspection reveals that several of the accounts generate suspiciously high volumes of partisan content.

\subsection{Topics from low-credibility sources}

\begin{figure}[t]
    \centering
    \includegraphics[width=0.75\columnwidth]{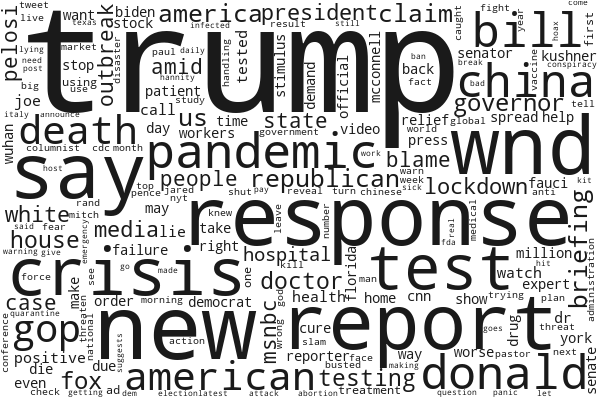}
    \caption{
    Word cloud of the most frequent words in the titles of articles in \texttt{DS2}.
    Font size indicates frequency.
    }
    \label{fig:wordcloud}
\end{figure}

We wish to provide a sense of the common topics of linked articles from low-credibility sources. Fig.~\ref{fig:wordcloud} depicts the most frequent words in the titles of the articles in \texttt{DS2}, excluding the query terms ``coronavirus'' and ``covid.'' 
Popular topics covered by low-credibility sources are U.S. politics, the status of the outbreak, and economic issues.

A closer analysis on low-credibility articles suggests a politicization of the pandemic.
An example revolves around claims that the COVID-19 pandemic originated from a weaponized virus. One of the most popular sources pushing this narratives is \textit{ZeroHedge.com}, the third most-shared low-credibility domain in our dataset. Notably, this occurs despite the fact that Twitter suspended the ZeroHedge account for violating the platform's manipulation policy at the beginning of the pandemic (\url{bloomberg.com/news/articles/2020-02-01/zero-hedge-permanently-suspended-from-twitter-for-harassment}).

\section{Discussion}

We characterize the prevalence of low-credibility information on Twitter during the novel coronavirus outbreak.
The combined prevalence of various low-credibility sources is comparable with mainstream and reliable sources.
Consistent with previous research, social bots are more likely to get involved in posting and amplifying low-credibility information. Finally, we find evidence of coordinated activity amplifying low-credibility content.

The analyses presented here are preliminary and have several limitations. 
The sampling method based on two hashtags might introduce unknown biases. 
The domain-based identification of low-credibility sources is not exhaustive and cannot capture misinformation contained in the content of tweets like text, images, and videos.
Bot detection algorithms are never perfectly accurate~\cite{cresci2017paradigm}. 
Our coordination analysis may be distorted by popular sources.
Finally, it is impossible to draw any conclusion about impact from the prevalence of misinformation alone.

Future work could confirm the robustness of our findings to alternative sampling methods, different definitions of low-credibility content, and additional methods to identify inauthentic accounts.
Granger-causality could be used to explore the impact of health misinformation on behaviors such as vaccination rates. It would also be interesting to study the characteristics of likely humans who spread low-credibility content about the pandemic.

\paragraph{Acknowledgement.} We are grateful to Pik-Mai Hui and Betsi Grabe for helpful discussions, and to DARPA (grant W911NF-17-C-0094), Craig Newmark Philanthropies, and Knight Foundation for support.

\fontsize{9pt}{10pt}
\selectfont
\bibliographystyle{aaai}
\bibliography{ref}
\end{document}